# Advanced scenarios for ITER operation

**A C C Sips**[1], for the Steady State Operation[2], and the Transport Physics[2] topical groups of the International Tokamak Physics Activity.

[1] *Max-Planck-Institut für Plasmaphysik, Boltzmannstrasse 2, D-85748, Garching, Germany*

[2] *For list of contributors see reference [54]*

E-mail address of the author: **ccs@ipp.mpg.de**

**Abstract**

In thermonuclear fusion research using magnetic confinement, the tokamak is the leading candidate for achieving conditions required for a reactor. An international experiment, ITER is proposed as the next essential and critical step on the path to demonstrating the scientific and technological feasibility of fusion energy. ITER is to produce and study plasmas dominated by self heating. This would give unique opportunities to explore, in reactor relevant conditions, the physics of α-particle heating, plasma turbulence and turbulent transport, stability limits to the plasma pressure and exhaust of power and particles. Important new results obtained in experiments, theory and modelling, enable an improved understanding of the physical processes occurring in tokamak plasmas and give enhanced confidence in ITER achieving its goals. In particular, progress has been made in research to raise the performance of tokamaks, aimed to extend the discharge pulse length towards steady-state operation (advanced scenarios). Standard tokamak discharges have a current density increasing monotonically towards the centre of the plasma. Advanced scenarios on the other hand use a modified current density profile. Different advanced scenarios range from (i) plasmas that sustain a central region with a flat current density profile (zero magnetic shear), capable of operating stationary at high plasma pressure, to (ii) discharges with an off axis maximum of the current density profile (reversed magnetic shear in the core), able to form internal transport barriers, to increase the confinement of the plasma. The physics of advanced tokamak discharges is described, together with an overview of recent results from different tokamak experiments. International collaboration between experiments aims to provide a better understanding, control and optimisation of these plasmas. The ability to explore advanced scenarios in ITER is very desirable, in order to verify the result obtained in experiments today and to demonstrate the potential to significantly increase the economic attractiveness of the tokamak.

**Introduction**

The ITER project [1] will provide a basis for the scientific and technological feasibility of a fusion power reactor. It aims to create for the first time, sustained deuterium/tritium plasma, predominantly heated by $\alpha$-particles produced by the fusion reactions ("burning" plasmas). It is a major step in the world fusion program, and a culmination of almost fifty years of magnetic confinement fusion research.

An essential feature of a tokamak plasma confinement scheme is the presence of a toroidal current in the plasma itself [2]. Normally, this current is established and maintained in the plasma by a transformer in the centre of the device (inductive drive). This implies that the configuration can be maintained only for a limited time, determined by the magnetic flux available for only cycle of the transformer coil. An increase in pulse length, or steady state operation, can be achieved if the toroidal plasma current is driven non-inductively. Heating and current-drive methods do exist to do this; the injection of energetic neutral atoms and powerful radio-frequency radiation. In addition, the plasma itself produces a non-inductive current associated with density and temperature gradients. This is the diffusion-driven, or bootstrap current [3].

ITER is a tokamak and represents an extrapolation of approximately a factor of 2 in linear dimension from the largest experiments today (see Table I). It will contain a plasma volume of more than 800 m$^3$ using a magnetic field of 5.3 T. ITER is designed as an experimental device with extensive diagnostics and a considerable flexibility in shaping, heating/current drive and fuelling methods. These are essential for accommodating uncertainty in projection, for exploring new operation regimes attractive for a reactor and for investigating new aspects of plasma physics, which may arise from e.g. significant $\alpha$-particle heating, large size and extended burn.

*The principal physics objectives of ITER [4] are:*
    (i)    To achieve extended burn using inductively driven plasmas with the ratio of

fusion power to auxiliary heating power ($Q$) of at least 10 for a range of operation scenarios and with a duration sufficient to achieve stationary conditions on the timescales characteristic of plasma processes.

(ii)     To aim at demonstrating steady-state operation using non-inductive current drive with a ratio of fusion power to input power for current drive of at least 5.

In addition, the possibility of higher $Q$ operation will be explored if favourable confinement conditions can be achieved. The rules and methodologies for the projection of plasma performance to the scale of ITER have been basically formulated in the ITER Physics Basis [1], which has been developed from broadly based experimental and modelling activities within the magnetic fusion programmes of the ITER parties.

It is also important to carry out engineering tests of components for future reactors, to test tritium breeding module concepts with the 14MeV neutron power load on the first wall $\geq$ 0.5MWm$^{-2}$ and fluence $\geq$ 0.3MWam$^2$. Operation of ITER is planned in two phases: An experimental physics-oriented program lasting 10 years and including, sequentially, operation in hydrogen, deuterium, and deuterium-tritium mixtures, followed by a 10 year long technology-oriented program. This second phase depends on reliable schemes for long pulse or steady state operation to be developed in the first 10 years of ITER operation. In this respect, ITER faces significant physics and technical challenges during its construction period and operation phases.

The reference scenario for ITER inductive operation is the H-mode [5], which has been observed in many tokamaks reliably and reproducibly. The properties of the H-mode have been investigated over the last 23 years, providing the basis for the achieving ITER's primary goal of operation at $Q=10$ based on scaling laws for projection. The energy confinement time is predicted using the IPB98(y,2) scaling law [1], while the average density in discharges should reach at least ~85% of the Greenwald density limit for H-mode operation $n_{GW} = 10^{20}$ $I_p[MA]/\pi a[m]^2$ [6], where $I_p$ (MA) is the plasma current, $a$ (m) is the plasma minor radius.

The standard ITER scenario does not allow conditions to be reached where the plasmas current is completely non-inductively driven, utilising the self generated bootstrap effect in the plasma. So called advanced scenarios in fusion experiments seek to improve confinement and stability over standard H-modes in order to maximise the bootstrap current. Important in tokamaks is the inverse rotational transform of a magnetic flux surface, the safety factor $q$.

Typically, for standard H-modes, $q$ in the centre ($q_0$) is the minimum $q$-value ($q_{min}$) and is just below 1, while $q$ at the plasma boundary ($q_{95}$, the safety factor at 95% of the plasma minor radius) is 3 or above. This safety factor $q$ and the magnetic shear $s=(r/q)dq/dr$ (with r the is the minor radius of the flux surface) play an important role in plasma stability and confinement. Key to the development of these scenarios is a tailoring and control of the current density profile. Advanced scenarios use a range of (non-monotonic) $q$-profiles as shown in Figure 1. It is of course possible to imagine a continuum of regimes between the reference non-inductive and inductive scenarios in which the current profile is modified externally but not completely driven by non-inductive means.

To date, two main types of advanced regimes are being developed. First, the "steady-state" advanced scenario, should provide the basis for satisfying ITER's second major goal of reaching $Q=5$ under fully non-inductive conditions. Typically these discharges have central $q$ above 1.5, with either weak $|q_0-q_{min}|$ ~0.5 or strong $q_0>>q_{min}$ reversed magnetic shear [7-12]. This kind of q-profile is used to obtain internal transport barriers (regions of reduced transport in the core), which could provide sufficient bootstrap current for steady state operation. A second advanced regime, the so-called "hybrid" scenario, has a stationary current density profile with weak or low magnetic shear, $q_0$~1 and $q_{95}$~4 [13-16]. This allows operation at high values for normalised beta, $\beta_N = <\beta>aB_T/I_p$ ($<\beta>$, volume averaged normalised pressure ($p$) in the tokamak $\beta = p/(B_T^2/2\mu_0)$, $B_T$ the toroidal field). Operating at lower plasma current compared to the ITER reference scenario, this regime could lengthen the discharge duration substantially (although not steady-state) and could play a key role in the second (technological) phase of ITER operation, in case the challenging requirements for full non inductive operation can not be met.

In the next sections of this paper, transport in tokamaks is described (section 2) as a physics basis for projection of the performance of the ITER reference scenario (section 3). This is followed, in section 4, by a description and overview of experiments with internal transport barriers. The hybrid scenario is presented in section 5. The results of an international collaboration activity to study and compare advanced scenarios are given in section 6, followed by an outlook for ITER and concepts for a fusion reactor (section 7). The results presented in this paper are summarised in section 8.

## 2. Transport and confinement in tokamaks

Understanding transport in magnetised plasmas is important for the design of future fusion reactors. A theory of the classical collisional transport losses has been developed. However, it does not completely explain the transport across magnetic surfaces. Hence, additional processes driven by plasma turbulence are required to understand the cross-field transport in tokamaks. This turbulence is mainly driven by two main micro-instabilities: Ion Temperature Gradient (ITG) driven modes and Trapped Electron Modes (TEM) [17,18]. In the non-linear regime these produce particle, momentum, electron and ion heat transport. The main characteristic of these micro-instabilities is the existence of an instability threshold (critical gradient); with a strong increase in turbulent transport above the threshold, although a finite diffusivity persists even when the gradient is below this critical gradient.

Over a significant part of the plasma cross-section, the logarithmic gradient of the temperature is close to the threshold; hence in this region the profiles are stiff. An exception to this is the edge of the plasma. Here energy transport across the plasma edge region is observed to decrease significantly when the input power is increased above a threshold value [2]. With this edge transport barrier (see Figure 2a), the total energy confinement of the plasma doubles (H-mode) over plasma without this edge transport barrier (L-mode). Access to the H-mode requires a configuration with a magnetic X-point so that the core plasma is separated from the plasma-wall interaction region, called divertor configuration. ITER is a divertor tokamak (see Figure 2b).

The energy confinement in a magnetic confinement device is characterized by a global energy confinement time $\tau_E = W/P$ where $W$ is the thermal energy stored in the plasma and $P$ is the input power to the plasma. Integrating a critical gradient model over the plasma volume could give a prediction for the core energy content. However, a correct prediction of the total energy content of the plasma will still depend, strongly, on the edge pedestal, and the degree of stiffness of the transport model. Despite results from experiments today, the interplay between electron and ion heat channels is unclear so no firm conclusion can yet be drawn regarding profile stiffness in ITER. Hence, the applicability of a critical gradient model for predictions to ITER has to be further investigated.

On the other hand, tokamaks with a range of sizes, operating parameters and heating powers have been constructed. Empirical scaling laws derived from confinement measurements of these experiments are useful for predicting plasma performance in any new device. A global scaling has been derived for the thermal energy confinement time of H-mode plasmas using the IPB98(y,2) scaling law [1].

$$\tau_{IPB98(y,2)} = 0.0562\ H_{H98(y,2)}\ I_p^{0.93}\ B_T^{0.15}\ P_n^{-0.69}\ <n_{e,19}>^{0.41}\ R_0^{1.97}\ M^{0.19}\ \kappa_a^{0.78}\ \varepsilon^{0.58} \qquad (1)$$

where $<n_{e,19}>$ ($10^{19}$ m$^{-3}$) is the line average electron density, $P_n$ (MW) the net heating power, calculated from the total input power, subtracting radiation losses [1], $R_0$ (m) denotes the major plasma radius, $M$ (amu) the average hydrogenic ion mass, $\kappa_a$ the plasma elongation defined as $\kappa_a = V/(2\pi^2 R_0 a^2)$ with $V$ being the plasma volume, and $\varepsilon = a/R_0$ is the plasma inverse aspect ratio.

The confinement of any H-mode plasma is then referred to this scaling using the confinement enhancement factor or H-factor ($H_{H98(y,2)}$), which is the ratio of the observed confinement to the scaling. Scaling laws are widely used to predict the energy confinement time in next step devices, and are a basis for comparing results from various different experiments today. Since no knowledge of the heating, temperature or density profiles, or atomic physics for that matter, is built into the analysis, a degree of uncertainty still exists in predicting the confinement properties and plasma performance (see next section). The energy confinement time predicted for ITER by the IPB98(y,2) scaling is 3.7s (for 40 MW additional heating, see Table I) with one technical standard deviation of ±14% and 95% log non-linear interval estimate of ±28%. The uncertainty for ITER predictions is not only statistical. When written in dimensionless form, the global scaling laws can still yield information on the mechanisms that underlie turbulent transport. For example, the log-linear form of the scaling law is equivalent to assuming that a single turbulence mechanism with one scale size is responsible for the transport. This seems unlikely to be the case for H-modes, where the core region may be dominated by short wavelength turbulence of the gyroBohm type, and the behaviour in the edge region is possibly determined by magneto hydrodynamic instabilities determining the height and width of the edge transport barrier These two processes will scale differently with the main scaling parameter $\rho*$, the normalized Larmor radius ($\rho* = \rho_s/a$, $\rho_s = (m_i T_s)^{1/2}/eB_T$ is the ion Larmor radius, $m_i$ is the ion mass and $T_s$ the temperature). Hence rewriting the IPB(y,2) scaling using dimensionless variables suggests that electromagnetic effects are important, either in turbulence itself or via plasma instabilities:

$$B_T\ \tau_{IPB98(y,2)} = (\rho*)^{-2.7}\ \beta^{-0.9}\ (\nu*)^{-0.01}\ q^{-3}, \qquad (2)$$

where, $v^*$ is the normalized collisionality defined as $v^* = v_{ei}qR/\varepsilon^{3/2}v_{Te}$ ($v_{ei}$ the electron-ion collision frequency, $v_{Te}$ is the thermal electron velocity). However, observation that H-mode confinement scaling is closer to the gyroBohm expectation, is confirmed by recent dedicated experiments in the JET and DIII-D devices [19,20].

In addition, improved confinement discharges described in this paper reveal that transport results from a balance between driving terms (gradients) and stabilising effects such as magnetic shear and velocity shear. A particular example are so called reversed magnetic shear discharges, discussed in section 4. Here a reversal of the magnetic shear in the core of the plasma allows formation of internal transport barriers. Not only useful for optimising performance of plasmas with the aim of obtaining fully non-inductive operation of a tokamak, but also important for understanding transport. Models for transport, not only need to be able to predict confinement in standard discharges, but also the existence of zones of reduced transport in the core of advanced scenario discharges, a hard challenge for any model for the turbulent transport in fusion plasmas. On the other hand, scenarios with improved confinement will increase the confidence that the ITER targets can be met, despite uncertainty in predicting the confinement of a device that is an significant extrapolation of results obtained today.

## 3. The ITER reference scenario

The H-mode is a reproducible and robust mode of tokamak operation with a long-pulse capability, and has been recommended as a reference scenario for inductive *Q~10* operation in ITER. Consolidation of this mode of operation is progressing well in experiments, with the aim of refining the fusion performance prediction and possibly finding ways to reach increased fusion power in ITER. Three main areas of research can be identified, and are summarised below: (i) operation at high plasma density, (ii) study and mitigation of Edge Localized Modes and (iii) the stability of the plasma.

*Operation at high plasmas density*
At plasma temperatures sufficient to sustain fusion reactions, it is advantageous to operate at high plasma density to ensure plasma purity together with good power and particle exhaust at the plasma periphery or divertor. However, density limits are observed in tokamak operation, mainly determined by the physics of the edge plasma. For H-modes, a maximum density for the sustainment of the edge barrier is observed, the Greenwald density limit. Typically, H-mode operation at densities approaching $n_{GW}$ is accompanied by a deterioration of the energy confinement. One of the major achievements in recent tokamak experiments [21] is the demonstration of H-modes with good energy confinement ($H_{H98(y,2)} \approx 1$) in plasmas with densities close to this Greenwald density.

*Study and mitigation of Edge Localized Modes*

In H-mode plasmas, so-called Edge Localised Modes (ELMs) periodically relax the edge pressure gradient (hence they are called ELMy H-modes). Moreover, stationary conditions can usually only be achieved with regular ELMs to control the particle behaviour. Most experiments observe so-called type I ELMs [22] with up to 10% of rapid (< 1ms) loss of stored energy in the plasma. As the ELM generates an energy pulse outward, this gives considerable concern for the lifetime of the plasma phasing components (the divertor) in a reactor scale device [23]. Extrapolating current ELM data to ITER, expected divertor target erosion rates are too excessive, although in recent estimates the expected ELM size is only a factor of 2 above the limit for target ablation (for carbon or tungsten targets) [24]. Hence, the mitigation of ELMs remains one of the priority research items for ITER.

*Stability of the plasma*

The safety factor $q$ plays a key role in plasma stability [1]. Conventional H-modes, have monotonically decreasing $q$-profiles with $q_0 < 1$. As a result, in the core of the plasma, periodic reconnections inside the $q=1$ surface, called sawtooth oscillations, flatten the pressure profile. Other magnetic (MHD) instabilities are observed at or near rational values of $q$. Neoclassical tearing modes (NTM's), occur at low-order rational surfaces (e.g. $q=3/2$), driven unstable by the local gradient of the equilibrium current density, and give a loss (10%-30%) of plasma stored energy. NTMs are often seeded by magnetic islands, resulting from a sawtooth collapse [25]. For the ITER reference scenario active means exist to stabilise, or reduce the impact of such modes; using electron cyclotron current-drive (ECCD).

*Prediction of ITER performance*

Sophisticated transport simulations codes are used for the calculation of the time evolution of plasma profiles (assumed toroidally axisymmetric) in a tokamak. Such plasma simulation codes are termed ´1.5D´ codes as the geometry of the magnetic surfaces is recomputed to be consistent with the detailed two-dimensional pressure balance of the plasma and magnetic field. The performance of ITER is predicted by means of a 1.5D transport code ASTRA [26], using transport coefficients based either on theoretical models or a prescribed radial dependence normalized to fit the scaling law for the thermal energy [27].

Certain reference scenarios [28] have been defined for design purposes, and assessed with these 1.5D transport codes in order to determine the 'envelope' of performance within ITER's capabilities. The 'operation' domain of one ITER scenario at 15 MA plasma current is given in Figure 3 where the fusion power is predicted for a range of the confinement enhancement $H_{H98(y,2)}$ in line with uncertainties in the energy confinement scaling. Curves for different values of $\beta_N$ are given, important for MHD stability and for the amount of fusion power (P$_{fusion}$) produced, as P$_{fusion}$ increases with $\beta^2$. A lower bound indicates a minimum amount of power required to sustain H-mode operation [29], together with an upper bound for the density *($<n_e>/n_{GW} = 1.0$)* in H-modes. This gives maximum and minimum fusion power predictions of 560MW and 260MW, respectively for $H_{H98(y,2)}=1.0$. The reference point for ITER is chosen at $H_{H98(y,2)} = 1$ and $<n_e>/n_{GW} = 0.85$, expecting 400 MW of fusion power at $Q = 10$ (see also Table I). This is at $\beta_N=1.8$, deemed save for avoiding excessive NTMs, that would otherwise significantly reduce (up to 30%) the energy confinement. As seen from Figure 3, about 7% of confinement margin is required to achieve operation with $Q = 10$ for $<n_e>/n_{GW} \leq 0.85$ (a save distance away from the density limit for H-modes). In this reference scenario, argon impurity dosing is used to keep the power flux to divertor region below 30 MW, which approximately corresponds to 5 MWm$^{-2}$ of target heat load, acceptable for current divertor target designs.

Although the realisation of the ITER working point has been demonstrated in experiments, optimisation continues, in particular in the three areas described at the beginning of this section. Part of this optimisation is also performed in research on the advanced scenarios presented in the remainder of this paper.

## 4. Scenarios with internal transport barriers

Obtaining stationary or steady state operation is key for advanced scenarios, but challenging as a tokamak maximises its fusion performance with inductive operation at high plasma current. Maintaining desired fusion performance ($P_{fusion} \sim 400$ MW for ITER) at lower plasma current implies foremost operation at higher $\beta_N$ compared to the ITER reference scenario ($\beta_N = 1.8$). In addition, operation at sufficient confinement ensures operation at $Q \geq 5$, allowing the use of input powers >40 MW to control the profiles in the plasma. In order to satisfy these objectives for the ITER non-inductive regime, several conditions have to be satisfied simultaneously. In particular, significant external current drive is foreseen together with the bootstrap current. lower hybrid current drive, with the highest current drive efficiency, energetic neutral beam injection and electron cyclotron current drive are used in today experiments. Since the bootstrap current is the consequence of local pressure gradients and proportional to $\beta_p$ ($\beta_p = 2\mu_0 <p>_a/<B_p>^2$ with $<p>_a$ the poloidal cross-section averaged plasma pressure and $<B_p>$ the average poloidal magnetic field on the plasma boundary), most experiments tend to operate at low current and with internal transport barriers (ITBs). The physics of ITBs is a broad subject that is already covered by several overview papers [30,12].

Important is that ITB formation has a power threshold; which is the amount of power that is necessary to produce a barrier. This is similar to the observation of a power threshold to produce an edge transport barrier for H-mode plasmas, but not necessarily governed by the same physics processes. Turbulent transport reduction due to $E \times B$ shear ($E$ is the local electric field in the plasma) flow is well documented [31,32]. Stabilisation results essentially from the shearing of turbulent convective cells, with two key ingredients playing a central role in the physics of the ITB formation: shear plasma flow and magnetic topology. Here, negative magnetic shear is known to decrease the drive for the turbulence, this effect is enhanced by the Shafranov shift of magnetic surfaces (also called $\alpha$ effect, $\alpha = -q^2 R(d\beta/dr)$ is a measure of the Shafranov shift). Note that the velocity shear rate will be small in a reactor at the onset of an ITB, so that magnetic shear and Shafranov shift will have to be optimised to trigger the internal transport barrier.

Reversed shear configurations are typically obtained by heating the plasma just after initiation, during the current ramp up phase with the current still diffusing in from the edge of the plasma. An increase in central temperature, using additional heating, slows down the current diffusion in the core, while the total current in the plasma is increasing towards the preset flat top value. This creates (transiently) an off-axis maximum for the current distribution in a tokamak. Additional current drive, a variation of the amount of central heating and changing the rate of rise of the plasma current allows different reversed shear configurations to be created. Once in these conditions enough additional heating is applied to allow the an internal transport barrier to form, a positive loop takes place where density and ion temperature gradients increase, thus boosting the velocity shear rate and allowing the confinement barrier to be sustained. Achieving stationary plasmas with internal transport barriers and $\beta_N \sim 3$ is a challenge. In particular as the non-linear interaction between the pressure profile and evolution of the current density profile determine the evolution of the ITB in time. In this section, experiments with a deep reversal of the magnetic shear are described which have so called "strong" internal transport barriers. This is followed by a presentation of results from experiments that optimise the stability of the plasma using "weak" transport barriers, including a demonstration for active control to maintain plasma profiles in ITB discharges close to their optimum shapes.

*Strong internal transport barriers*

Strongly reversed shear configurations have been studied in view of ITER steady state operation. Typical ITBs in reversed shear configurations are found to be located at the vicinity of the location of $q_{min}$. Once such strong ITBs are formed, the bootstrap current is driven locally at the transport barrier location, and the shear reversal becomes larger. In an extreme case, central toroidal current can be almost or absolutely zero, hence the name of current hole given to these particular scenarios [33] Within the operation boundaries for stable operation, these discharges achieve bootstrap fractions of ~50%. In order for the shear reversal to be maintained, a certain amount of off-axis current drive is required, driving ~ 50% of the plasma current in these strongly reversed shear plasmas. However, the steep pressure gradients in these plasmas tends to create MHD instabilities, such that the required beta for operation with dominant bootstrap fraction can not be achieved. An example of a reversed shear discharge, with a current hole in the centre, is given in Figure 4. High transient performance is obtained; using electron cyclotron current drive and preheating with neutral beam under feedback control to obtain a stable current rise phase [33]. This heating scheme gives reproducibility for these types of discharges with enhanced confinement and performance of the plasma. The performance phase ends with a global instability terminating the discharge (called disruption). This is an ideal kink instability ($n=1$, with n the toroidal mode number of the instability) of the plasma column, driven by the steep local pressure gradients associated with the strong internal transport barrier (Figure 4c) [34]. Typically, stationary operation with strong reversed shear is at low beta (see as well Figure 10a in section 6, which using an international data base gives a comparison of all experiments). This is also the case for stationary reversed shear discharges in JET using lower hybrid current drive, even with control of the ITB strength to maintain appropriate current density and pressure profiles [35] in these experiments, $\beta_N$ is limited to values below 2.

In addition, discharges with strong ITBs do observe an accumulation of plasma impurities in the core, diluting the fusion fuel mix [36]. Moreover, the fast-ion confinement in current-hole plasmas was studied JET using tritium neutral beam injection in deuterium plasmas [37]. The confinement of alpha particles was determined from the decay time of γ-ray emission after tritium beam switch-off, with the γ-ray emission coming from the reaction of the fusion alphas and beryllium impurities (In JET beryllium is used to getter oxygen in the vacuum chamber). Results, indicate a α-confinement degradation of a factor ~5 in strongly reversed shear discharges, compared to conventional H-modes, as predicted from orbit losses by 3-D Fokker Planck codes. Both the observation of impurity accumulation and prompt alpha losses would be unacceptable for obtaining a stable burning plasma in ITER.

*Weak internal transport barriers*

Discharges with reversed magnetic shear can gain in stability against ideal kink modes using weaker ITBs with broader pressure profiles [34]. One route is to produce ITBs at larger radii (away from the plasma centre). This can be done by operating at plasma currents such that $q_{95}$~3, creating a wide region with reversed magnetic shear by heating during the current rise phase, or at much lower plasma current using off-axis current drive. It is observed that these transport barriers are often positioned close to a low order rational surface in the vicinity of the plasma boundary [38] (near *q=2* for the high current discharges, near *q=3 or 4* for the low current cases). These transport barriers have weaker density and temperature gradients improving stability and show no signs of accumulation of impurities in the core. So far however, these plasmas have not achieved performance levels required for ITER advanced scenarios, the main difficulty being the simultaneous optimisation of the edge conditions (avoiding type I ELMs, which can erode the internal transport barriers) and *q*-profile to maintain the barrier.

Maintaining, even these weakened ITBs for durations longer compared to typical energy and current diffusion time scales is a challenge. Multi-variable, model-based, techniques have been developed [39] for the real-time control of the current profile and/or the pressure profile, to ensure stationary conditions and MHD stability of the discharge. In experiments in JET, a first successful demonstration of combined electron temperature and current density profile control in advanced tokamak regimes has been obtained. Closed feedback loops using three actuators, the input power from lower hybrid current drive, ion cyclotron resonance heating and neutral beam injection systems, were used in conditions reaching up to 100% non-inductive current drive [40]. However, these results are still in ITBs plasma with a rather modest plasma performance ($\beta_N < 2$) using discharges at low plasma current ($q_{95} \geq 7$), to maximise the bootstrap faction and externally driven non-inductive fractions. This is similar to results obtained in JT-60U, where in discharges with a weak ITB and broad pressure profile produce up to 80% bootstrap fraction even though, $\beta_N < 2.2$ in such discharges. Due to lower $\beta_N$ limit, these experiments have been performed at $q_{95} \sim 9$ in JT-60U, in order to attain high enough $\beta_p$ [41].

On the other hand, discharges with weak magnetic shear ($|q_0-q_{min}| \sim 0.5$) also produce less strong barriers. In this cases the moderately peaked pressure profile, prevents the bootstrap current from peaking off-axis, and the shear from reversing too strongly; sustaining a rather flat $q$ profile with $q_0 > 1.5$ in the core. An example of such regime is given in Figure 5 [42] showing a typical time evolution of a weak reversed shear discharges in the DIII-D tokamak. Here, central neutral beam current drive together with off-axis electron cyclotron current drive is used together with the bootstrap current to create the desired weak transport barriers. These discharges start with an H–mode induced early in the current ramp. During the high performance phase, $\beta_N = 3.1$ is maintained by feedback control of the neutral beam power. Approximately 2.5 MW of co-directed ECCD resonant off-axis at 40% of the minor plasma radius is applied starting at 3.0 s. Between 3.0 and 4.0 s, the current density profile is observed to be nearly constant with $q_0 \sim 2.1$ and $q_{min} \sim 1.7$. The total non-inductive current drive in this case approaches 95%, with 65% of the plasma current provided by bootstrap current, 20% by neutral beam current drive. These discharges typically achieve $\beta_N$ up to *3.5*, for several energy confinement times. The termination of these conditions is due to the resistive evolution of the current profile, leading to the onset of NTMs as $q_{min}$ crosses 1.5. This results support the observation that the loop voltage profile is not fully relaxed, i.e. the net Ohmic current is almost zero, but the local Ohmic current is not zero everywhere. Hence, this regime requires further optimisation to obtain stationary conditions.

Along the lines of reducing the negative shear in the centre, it has been observed that ITBs can also been formed in plasmas with low magnetic shear. The so-called high $\beta_p$ plasmas in JT-60U belong to this category [43]. A series of full non-inductive current drive experiments at high performance have been achieved in this way. In these high $\beta_p$ plasmas, magnetic shear is low or even positive and $q$ profiles can vary from those with $q_0$ slightly in excess of 1 to those with $q_0$ around 2. Steady state demonstration discharges have been obtained with $q_0$ is below 1.5 [44]. As for the DIII-D discharges described above, on-axis neutral beam heating is a key feature to achieve full current drive conditions. In JT-60U neutral beam injection at an energy of ~350 keV is used, (compared to typically 80-120 keV in other experiments), from negative ion based sources, to increase the current drive efficiency.

The next section goes even a step further were an intermediate, or Hybrid, between the conventional scenario and the weak reversed shear scenarios is given with *1.0 ≤ $q_0$ ≤ 1.5*. Consequently, the JT-60U results with *$q_0$ < 1.5* (described above) are now assigned to this regime.

## 5. Hybrid scenarios

Scenarios with internal transport barriers, presented in the previous section, rely on a careful tailoring of the current density profile by external heating and current drive methods, to optimise performance. This way, sufficient bootstrap current may be provided to satisfy ITER's second major goal of reaching $Q \geq 5$ under fully non-inductive conditions. However, the stringent control requirements for scenarios with internal transport barriers have prompted research into advanced regimes, which are inherently stationary with respect to the current relaxation time scale, requiring only minimum control by external actuators.

It was originally envisioned [1] that discharges with extended duration at lower plasma current would be intermediate between an inductive (baseline) scenario and a fully non-inductive (advanced steady state) scenario. Therefore, this type of discharge is known as a "hybrid" scenario. This will allow ITER to operate in a mode maximising the neutron fluence for the purpose of testing the design of various components (second operation phase of the project). It has been found that scenarios with a stationary current density profile, maintaining zero magnetic shear in the centre permit to achieve such a target. The different $q$-profile of the hybrid scenario, compared to the standard inductive H-mode scenario, prevents sawtoothing activity in the core and the triggering of large neo-classically tearing modes at the *q=3/2* rational surface. These MHD events generally lead to significant reduction in confinement and limit plasma performance for *$\beta_N$ > 2* as observed in the standard H-mode regime with $q_{95}\sim 3$. The properties of the current density profile allow the hybrid scenario to operate at $\beta_N\sim 3$, suggesting that it could even provide an alternative route to establishing $Q=10$ in ITER [2]. Rapid progress has been made recently in developing this regime, and is described below.

In 1998, the ASDEX Upgrade tokamak found a stationary regime with improved core confinement for both electrons and ions in combination with an H-mode edge [13]. Initially, the pressure increase in the core was attributed to the formation of an internal transport barrier. However, quickly after, detailed transport analyses showed that in such a regime the temperature profiles remain in the so-called stiff regime. The gradients do not exceed a critical temperature gradient length set by the turbulence in the plasmas, and hence, no ITB is produced [45]. This new regime was called "Improved H-Mode". Further development of these type of discharges in ASDEX Upgrade [46,47] and DIII-D [48,49] are now known under the common name "ITER Hybrid Scenario". The desired $q$-profile is obtained by heating during the current rise phase of the discharge, at moderate neutral beam power (2.5 to 5 MW). In the subsequent main heating phase, beta can be increased, with either small MHD $m=1/n=1$ activity in the core (called fishbone activity, with m the poloidal mode number of the instability) or a small $m=3/n=2$ neoclassical tearing modes creating a central $q$-profile with very low magnetic shear and $q_0$ near 1. These discharges have no sawteeth and peaked density profiles with $H_{H98(y,2)}$ up to 1.4 for the duration of the heating phase. An example of an ASDEX Upgrade discharge is given in Figure 6. Note that increasing the neutral beam power after 3 seconds in the discharge leads to a strong rise in beta to $\beta_N\sim 3$, together with a improvement in confinement. Typically, these discharges obtain non-inductive current fractions of ~50%, in combination with benign MHD modes in the core, maintaining a stationary $q$-profile without active control.

Recently, experiments in JET establish the hybrid scenario in similar non-dimensional parameters (for example: $\rho^*$ and $q$-profile, $q_{95}\sim 4$) compared to ASDEX Upgrade [50,51]. These discharges are stationary for the duration of the heating phase with small NTM and fishbone activity in the core at similar $\beta_N$, $H_{H98(y,2)}$-factor, density and temperature profiles compared to ASDEX Upgrade or DIII-D.

Further evidence that the hybrid scenario may be a natural operating point for a tokamak comes from experiments in JT-60U. As described in the previous section, stationary high performance is obtained in the so-called "high $\beta_p$ ELMy H mode regime" [43], closely resembling the conditions obtained in other experiments. Discharges without sawteeth have been sustained at $\beta_N = 2.6$ for 2.6 s with $q_{95}\sim3.4$ [44]. More recently, discharges have been obtained with longer duration (see next section) at somewhat lower beta, $\beta_N=3$ was sustained for a shorter time (6s). These recent experiments confirm the potential of the hybrid scenario to reproducibly obtain improved confinement and stability over standard H-modes in stationary conditions.

For extrapolation to ITER, a mapping of the operational domain of the hybrid scenario at various plasma densities has started, in a collaboration between various experiments. ASDEX Upgrade demonstrated operation of this regime at 80% to 90% of the Greenwald density limit. Discharges with a high triangularity configuration of the plasma cross-section, $\delta=0.43$, showed only a small reduction in confinement ($H_{98(y,2)}=1.1-1.2$) compared to hybrid discharges at lower density, while sustaining $\beta_N=3.5$ and peaked density profiles. Moreover, the type I ELM activity in these discharges is moderated and small amplitude (called type II) ELMs are observed when the discharge configuration is changed to have a double null divertor configuration (both an upper and lower divertor are active, termed "double null" configuration) [52]. This is shown in Figure 7, where the heat load to the lower divertor is measured with infrared camera diagnostics. Initially, large transient heat loads (> 18MW/m$^2$) are observed, during the type I ELMing phase (as described before, a concern for ITER). As the plasma shifted to a double null configuration at 3.5 seconds, smaller type II ELMs appear. During this phase the outer divertor shows a near continuous power load of < 6 MW/m$^2$, the inner divertor has no power load as part of the power now goes to the top of the device. These type II ELMs, when extrapolated to ITER would have acceptable power loading of the ITER divertor target. These small ELMs can also be obtained in conventional H-modes, although with $H_{H98(y,2)} < 1$ [52]. The hybrid scenario compensates for the confinement loss with improved core confinement, achieving $H_{H98(y,2)} \sim 1.1$. However, the precise nature of the improved confinement in Hybrid scenarios is under investigation.

## 6. The international tokamak physics activity for advanced scenarios

While the inductive H-mode is relatively well explored, an open issue is how the presently developed advanced scenarios will extrapolate to next-step experiments. The scientific progress in preparation of ITER now benefits from a coordinated experimentation between tokamaks worldwide, in particular between large and middle-size D-shape tokamaks such as JET, JT60-U, DIII-D or ASDEX-Upgrade. The past two years have seen an unprecedented number of similarity and identity experiments involving two or more tokamaks, as well as coordinated parametric scans. These experiments are initiated by the International Tokamak Physics Activity (ITPA).

Construction of an international database for advanced tokamak discharges is also an activity coordinated under the ITPA [53]. Data from ASDEX Upgrade, DIII-D, FT-U, JET, JT-60U, RTP, T-10, TCV, TFTR and Tore Supra experiments have been collected, creating a set of scalar data covering a wide range of plasma parameters. Extensive analyses of this database has been presented before, using values at the time of maximum performance during the discharges [54]. This has recently been extended to document the differences in operational domain of reversed shear and hybrid scenarios, taking data as an average over the performance phase, as this is more appropriate for scenarios developed for stationary or non-inductive operation.

In the analysis of the data a figure of merit defied as $H_{89}\beta_N/q_{95}^2$ is used for performance: $H_{89}\beta_N/q_{95}^2 \sim 0.40$ for the ITER reference scenario, and $H_{89}\beta_N/q_{95}^2 \sim 0.3$ for the ITER non-inductive scenario. The confinement enhancement factor, $H_{89}$, relative to *ITER89P* scaling [1] is used, as this is more suited for a dataset containing discharges with a variety of edge conditions (L-mode, ELM free and H-modes with various types of ELM behaviour). The operation space and the performance of the advanced scenarios described in the previous two sections are compared with data from the different experiments.

*Reversed shear scenarios*
The results for reversed shear scenarios with internal transport are presented in Figure 8, showing the performance from several machines plotted as function of the duration of the discharges, normalised to the energy confinement time. Transient discharges (duration < $10\tau_E$) can obtain performance exceeding ITER requirements, but this cannot be maintained at these levels in more stationary conditions (duration $\geq 10\tau_E$). The reversed shear discharges separate into two distinct groups, dominated by data from DIII-D on the one hand, and data from JET and JT-60U both at lower performance. ITER, with a inductive discharge duration of 400 seconds and a energy confinement time predicted to be 3.7 seconds, would have a normalised duration of 110 is outside the range used for plotting the results from reversed shear scenarios. The duration of the experiments is mainly determined by machine limits, i.e. duration of the heating systems. This implies that reversed shear discharges that are stationary, with respect to the current diffusion time, typically $(30\text{-}50)\tau_E$, and have a duration longer compared to the ITER conventional scenario ($\sim 110\, \tau_E$) still need to be demonstrated.

More results are given in Figure 8b where $H_{89}\beta_N/q_{95}^2$ is plotted versus $\varepsilon^{0.5}\beta_p$. The latter is a measure for the fraction of the bootstrap current for similar $q$-profiles; such as the reversed shear discharges plotted in this figure. For reference, detailed transport analyses, including an assessment of the non-inductive current distribution of a few discharges in this database, indicate that for $\varepsilon^{0.5}\beta_p = 1$, a bootstrap fraction in the range 55%-65% is achieved. Internal transport barrier discharges with $q_{95}<5$ are only transient. Stationary operation has been obtained at $q_{95} \geq 5$, with the maximum performance (discharges from DIII-D) obtaining $\varepsilon^{0.5}\beta_p \sim 1$ for $q_{95}$ near 5. With $H_{89}\beta_N/q_{95}^2 \sim 0.3$ and sufficient bootstrap current, combined with external current drive sources, these discharges fulfil the ITER requirements for non-inductive current operation at $Q\sim5$. The results at $q_{95} \geq 6$ are from discharges with large, weaker ITBs predominantly operating in stationary conditions at low plasma current. Despite producing bootstrap current fractions in the range 40%-80% or using real time control techniques to optimise the profiles, they fail to meet the ITER performance targets, as either the energy confinement time or the achieved beta values are too low to ensure sufficient fusion power.

*Hybrid discharges*

The duration of hybrid discharges is typically longer compared to reversed shear plasmas. This is shown in Figure 9a (note that the time axis in Figure 9a is different from the axis used in Figure 8a). There is no clear difference between the various experiments in the dataset (only the Tore Supra data, have lower performance), and typical performance of hybrid scenarios achieves ITER reference values for $Q\sim10$ operation or higher. The duration approaches ITER target values for the conventional scenario, limited again by machine hardware. For high fluence operation in ITER a demonstration of longer duration pulses is required, although difficult to obtain: For example, discharges in JT-60U have obtained $\beta_N \sim 1.9$ for 24 seconds for $q_{95}=3.3$, with the $q$-profile matching hybrid conditions, obtaining $H_{89}\beta_N/q_{95}^2=0.40$ for about $120\,\tau_E$ (not shown in Figure 9a). Here the long duration is achieved by applying the different beam sources in subsequent phases (hence not at maximum input power or beta) to compensate for the limited duration of the neutral beam heating systems (typically 10 seconds) in these experiments. Most of the highest performance pulses shown, are a direct result of collaboration between experiments, benefiting from an exchange of expertise to optimise the hybrid regime. Such a cross fertilisation between experiments has not yet taken place for reversed shear scenarios with internal transport barrier as seen from the separation of the results from various devices in figure 8a.

The fraction of the bootstrap current in hybrid discharges is lower, compared to reversed shear discharges, and increases at lower current (higher $q_{95}$) as shown in Figure 9b. For this type of $q$-profile with $q_0$ near 1, a value of $\varepsilon^{0.5}\beta_p.=1$, achieved for discharges at $q_{95} = 4$-$4.5$, represents a bootstrap fraction of 30%-40%. This implies that this regime can only be used for long pulse operation with a substantial increase in discharge duration (see next session). On the other hand, at lower values of $q_{95} \sim 3$, the results of hybrid discharges far exceed ITER performance targets for operation at $Q=10$. Operating with $H_{H98(y,2)}$ in the range *1.0-1.3* and $\beta_N$ values between *2.5-3*, these discharges would allow a significant increase in fusion power for ITER, as can be seen from Figure 3. In the conventional scenario, operation at $q_{95}$ below *3* significantly increases this risk of sawteeth instabilities and NTM modes. Now hybrid scenarios are being develop with $q_{95}<3$, without sawteeth and low magnetic shear in the core, as this mode of operation may achieve stationary conditions at $\beta_N > 2$ to allow for controlled ignition ($Q > 20$) experiments in ITER at 17 MA.

*Beta limits*
Advanced scenarios maximising the fraction of self-generated bootstrap current, are likely to operate near one or more stability limits. In general, discharges can gain in stability against ideal $n = 1$ kink modes by optimising the pressure profile and plasma shape [34]. Kink modes can manifest themselves as resistive wall modes [55], which set in when the plasma pressure typically exceeds $\beta_N \sim 4li$ (with *li*, the inductance of the plasma). Reversed shear discharges, have low plasma inductance ($li < 0.8$), as the current density peaks off axis, and have predominantly peaked pressure profile due to the presence of (weak) internal transport barriers. Hence, they are at a particular disadvantage with regard to kink stability. Figure 10 plots the $\beta_N$ values achieved in the advanced discharges of the ITPA database as function of the pressure peaking ($p_0/<p>$, calculated using central density, central temperatures and plasma stored energy). The data support previous studies that the maximum $\beta_N$ drops sharply for high pressure peaking [56], in fact 80 % of the reversed shear discharges in the database achieve the high performance only transiently, limited to $\beta_N < 2$. Discharges with strong reversed shear, achieve a pressure peaking, $p_0/<p> >4$, and have with very good transient confinement ($H_{H98(y,2)}$ ~*1.5*) compared to conventional H-modes. In reactor plasmas with $P_\alpha > P_{input}$ and sufficient confinement, the fusion performance is set by the ability of the plasma to operate at the maximum plasma pressure possible ($P_{fusion} \propto \beta^2$). Hence, reversed shear discharges with a low beta limit are not useful for fusion reactors. On the other hand a promising scenario uses weak negative magnetic shear, with broad pressure profiles produced by weak transport barriers. Here results from DIII-D and JT-60U show that $\beta_N$ ~ *3* can be obtained with nearly 100% non-inductive current. This suggests an optimum in the range of *q*-profiles suitable for advanced operation, with only weak negative shear and $q_0$ =*1.5-2*, not far away from *q*-profiles used in the hybrid scenario. Figure 10b indicates that these hybrid scenarios operating with $q_{95}$ in the range *3* to *4.5*, routinely obtain beta values close to the no wall limit *($\beta_N/4l_i$~1)*.

Building on the success of the experiments coordinated under the ITPA, further study and collaboration between experiments are being defined. These include (i) a continuation of the documentation of the operational space for hybrid discharges, planning experiments to optimise the regime at lower $\rho^*$ in JET and JT-60U and further documentation of the regime at ASDEX Upgrade and DIII-D, (ii) an effort to define a common reversed shear scenario for various devices to, in particular concentrating on operation near $q_{95}$=*5*, (iii) perform transport studies in a range of scenarios as, for example, the improvement in confinement for the hybrid scenario is not fully understood, and (iv) operation in a parameter range closer to ITER values to facilitate a more robust extrapolation of the results obtained; for instance new experiments to confirm, that injected momentum from the neutral beams is not essential, as found in preliminary experiments in ASDEX Upgrade [57] and JET [58] using ICRH heating, to simulate the heating conditions in ITER or a reactor.

7. **Predictions for ITER and outlook**

ITER has a flexible design, capable of exploring the advanced scenarios presented in this paper. It could incorporate the use of current drive methods, such as neutral beams, ion cyclotron, electron cyclotron, and lower hybrid waves are important for reversed shear scenarios. In the ITER plans, operation will start with a total additional power of 73 MW: A neutral beam system will provide 33 MW in atomic deuterium beams at 1 MeV from two injectors, with the capability of providing on-axis and near off-axis current drive. An additional 40 MW will be available from radio frequency heating and current drive systems. In using this as boundary conditions, performance predictions for the various scenarios for ITER have been made using 1.5D transport simulation codes [59]. The conventional scenario given in section 3 is summarised in Table I, together with predictions for two advanced scenarios.

As a candidate scenarios for steady-state operation a 'weak' negative shear configuration is chosen, requiring off axis current drive using the neutral beam system. Absence of *q = 2, 1.5* and *1* surfaces inside the plasma eliminates the growth of tearing modes. Although the normalised beta and $H_{H98(y,2)}$ are relatively high, the ITER PF-coil system is designed to cover this operation scenario. In the simulations, realistic neutral beam heating and current drive modelling in ITER geometry has been used. For the radio frequency heating and current drive systems, the absorption and current drive efficiency ($\gamma_{20}$ = *0.3 AW$^{-1}$m$^{-2}$*) are predicted for ITER plasma conditions. The weak negative shear scenario still uses substantial external current drive (50% of the total current). In this context, one needs to bear in mind that in a steady state tokamak reactor, a fraction of the bootstrap current of ≥70% is required for economical operation.

1.5D transport simulations are performed for the hybrid scenario. The results show that this scenario already satisfies operation at *Q ~ 5* for $H_{H98(y,2)}$ = *1.0.* with a burn time ≥ 1000 s assuming a start of the additional heating during current ramp up phase [55]. The improved energy confinement found in recent experiments, if realized in ITER, could significantly improve the plasma performance in the hybrid scenario allowing *Q ≥ 10* at reduced plasma current (Table I). This simulation only uses 30 MW neutral beam power for the hybrid scenario. However, this scenario is capable of operation at $\beta_N$ *~3,* which would allow the full 73 MW of input power to be used, providing a substantial increase in fusion power and an increase in discharge length. Future developments of hybrid scenarios might even lead to the concept of "quasi steady state" reactor. A system with minimal time between two subsequent very long duration discharges, aiming at a duty cycle of > 90%. This would reduce the corresponding thermal stresses, which are seen as the main limitation of such a pulsed tokamak reactor proposal.

*One step to a demonstration fusion reactor ?*
ITER is expected to play an important role in the fusion development strategy: one step to a reactor producing electricity. A number of design studies have been carried out on fusion power reactors. In a recent European Power Plant Conceptual Study, four conceptual designs for commercial fusion power plants were presented each with a net electrical output chosen to be around 1500 MW electric power [60]. The model designs (A,B,C and D) differ in their dimensions, gross power, and power density. Models A an d B have the largest plasma dimensions (model A: $R_0$=9.55m, $a$=3.2m) and are based on about thirty percent improvements ($H_{H98(y,2)}$=1.2, $\beta_N$=3.5, 45% bootstrap fraction) on the design basis of ITER. This is in line with the performance assumed for the Hybrid scenario. Even with these assumptions, the devices have a large system size and high plasma current (model A: 30.5 MA). Figure 11 shows the plasma cross sections of models compared to ITER. The technology employed in models A and B stems from the use of near-term choices. The other designs (C and D) are based on progressive improvements in the level of assumed development in plasma physics, especially in relation to plasma shaping and stability, limiting density, and in minimisation of the divertor loads without penalising the core plasma conditions. Hence, they represent possible future improvements of the tokamak concept leading to more efficient and economical reactors. Clear is that in near term the results of ITER advanced scenarios are important, to make it a one step to a demonstration reactor. For example, as these four conceptual designs are steady state reactors, the efficiency assumed for the current drive system (250 MW for model A), need to be verified by operating ITER at the highest, reactor relevant, temperatures.

## 8. Conclusions

ITER, is an essential step to develop the scientific and technical feasibility of fusion energy. It is based on a magnetic confinement concept called the tokamak; an inherently pulsed system where a current in the plasma plays an important role in the confinement and stability if the system. Magnetically confined plasmas at finite pressure have turbulent transport across the magnetic field lines, driven by temperature gradients. Hence, predominantly empirical scaling and system modelling is used to predict the energy confinement and system size of ITER so its primary goal for obtaining conditions with significant fusion power and gain $Q\geq10$ can be met. Stabilisation, or reduction of the turbulence allows transport barriers to be formed, improving overall confinement. Important for creating and sustaining transport barriers is the local magnetic shear in the plasma. Discharges with improved confinement, employing internal transport barriers are called advanced scenarios. These are being developed to meet ITER's second goal of steady state (non-inductive) operation with $Q \gtrsim 5$, allowing for a reduced gain to operate at lower plasma current, and to use additional heating to control the current density profile. To achieve this, the performance of tokamaks needs to be improved. Standard tokamak discharges have a current density increasing monotonically towards the centre of the plasma. Reversed shear discharges have an off axis maximum of the current density, creating negative magnetic shear in the core. Results show that the confinement is indeed improved in plasmas with internal transport barriers. However, the plasma pressure required to obtain sufficient self generated, bootstrap current and fusion power, can not be sustained due to MHD stability limits. Only discharges with weak reversed shear maintaining a reduced internal barrier strength approach the required performance. This has prompted research into so called hybrid regimes that have a central region with a flat current density profile (zero magnetic shear), capable of operating stationary at high plasma pressure. Recently, international collaboration between experiments has enabled better documentation of the various types of advanced scenarios used and an exchange of expertise to optimise these regimes. The optimum advanced scenario operates with a current density profile which is close to the non-inductive scenario, but significantly different to allow an increase in stability and (some) reduction of the turbulent transport. As a result, recent tokamak fusion reactor concept studies using conservative extrapolations, require a large system size (1.5 times ITER), and a substantial additional heating to drive ~50% of the plasma current non-inductively. In summary, it is critical to study advanced scenarios in ITER to increase physics understanding of transport and stability of fusion plasmas, and secondly to validate and improve these regimes in preparation for a fusion power plant demonstration.

**Table I:** Predicted performance for three ITER scenarios

| *Parameters* | *Reference* | *Reversed shear[1]* | *Hybrid* |
|---|---|---|---|
| Major radius, $R_0$ (m) | 6.2 | 6.35 | 6.2 |
| Minor radius, $a$ (m) | 2.0 | 1.85 | 2.0 |
| Toroidal field at $R_0$, $B_T$ (T) | 5.3 | 5.2 | 5.3 |
| Plasma current, $I_p$ (MA) | 15 | 9 | 12 |
| Edge safety factor, $q_{95}$ | 3.0 | 5.3 | 4.1 |
| Confinement enhancement, $H_{H98(y,2)}$ | 1.0 | 1.57 | 1.2 |
| Normalised beta, $\beta_N$ | 1.8 | 2.95 | 2.1[*] |
| Average electron density, $<n_e>$ ($10^{19}$m$^{-3}$) | 10.1 | 6.7 | 8 |
| Fraction of Greenwald limit, $<n_e>/n_{GW}$ | 0.85 | 0.82 | 0.85 |
| Average ion temperature, $<T_i>$ (keV) | 8.0 | 12.5 | 8.8 |
| Average electron temperature, $<T_e>$ (keV) | 8.8 | 12.3 | 9.9 |
| Neutral beam power, $P_{NB}$ (MW) | 33 | 33 | 30 |
| RF power, $P_{RF}$ (MW) | 7 | 29 | 0 |
| Fusion power, $P_{fusion}$ (MW) | 400 | 356 | 367 |
| Fusion gain, $Q=P_{fusion}/(P_{NB}+P_{RF})$ | 10 | 6 | 11 |
| Non inductive current fraction, $I_{NI}/I_p$ (%) | 28 | 100 | 44 |
| Burn time (s) | 400 | 3000[§] | 1550 |

[1] This scenario has $q_0=3.5$ and $q_{min}= 2.2$, and uses a plasma configuration that is shifted outwards.
[*] Could go to higher normalised beta with increased input power.
[§] Limited by ITER plant restrictions.

**Figure Captions**

**Figure 1:** The range of $q$-profiles for the ITER conventional scenario (blue), the Hybrid scenario (green) and the reversed shear scenario (orange), showing weak and strongly reversed scenarios.

**Figure 2**: A schematic representation of a H-mode plasma with and an edge pedestal (a). The core and pedestal regions are also indicated on an ITER plasma cross section (b). Also shown is the divertor region, used for power exhaust.

**Figure 3:** A simulation of the operation domain in the $H_{H98(y,2)}$-factor and fusion power space when $I_p$ = 15MA and $Q$ = 10 [59].

**Figure 4:** a) Waveforms of a high-performance reversed shear (current hole) discharge in JT-60U [33], showing the increase in plasma current and applied neutral beam power (top traces), the electron temperature ($T_{e0}$) and ion temperatures in the center ($T_{i0}$), together with the electron cyclotron heating power($P_{EC}$) (middle traces) and the rise of plasma stored energy (W) and the D-D neutron yield (bottom traces) (b) $q$-profile at 7.2 s and (c) the evolution of $T_i$ during the high performance phase and $T_e$ profile at 7.2 s.

**Figure 5:** Plasma parameters versus time for a discharge in DIII-D at high beta, in which off axis co-ECCD is used to maintain the current profile: (a) plasma current (MA), neutral beam injected power (10 MW), line-averaged density ($10^{20}$ m$^{-3}$), (b) $\beta_N$ (black trace), $4l_i$ (green trace) and ECCD power (a.u.), (c) $q_0$ (upper trace), $q_{min}$ (lower trace), (d) central ion and electron temperature [42].**Figure 6**: Waveforms of a hybrid discharge at ASDEX Upgrade. Shown are (a) the plasma current (MA) and $D_\alpha$ measurements in the divertor (showing the ELM behaviour), (b) the neutral beam power (MW) applied, (c) the plasma inductance and the normalised beta ($\beta_N$), and (d) the confinement enhancement factor $H_{H98(y,2)}$ and averaged electron density ($<n_e>$) normalised to the Greenwald density limit ($n_{GW}$).

**Figure 7:** Example of a hybrid discharge at high density ($<n_e>/n_{GW}$ ~0.85) from ASDEX Upgrade. (a) Plasma current and the and $D_\alpha$ measurements in the divertor (showing small ELMs). (b) The infrared measurements of the power loads on the inner lower divertor. (c) The infrared measurements of the power loads on the outer lower divertor.

**Figure 8:**     Data from the ITPA database for reversed shear discharges from various devices (colour coded). (a) The performance ($H_{89}\beta_N/q_{95}^2$) as function of discharge duration, closed symbols are transient discharges, closed symbols are stationary, the duration ($W>0.85$ maximum $W$) is normalised to the energy confinement time ($\tau_E$) averaged during this time window. (b) $H_{89}\beta_N/q_{95}^2$ versus $\varepsilon^{0.5}\beta_p$. The lines indicates the different values for $q_{95}$, transient (open symbols) and stationary results (closed symbols) are given.

**Figure 9:**     Data from the ITPA database for hybrid shear discharges from various devices (colour coded). (a) The performance ($H_{89}\beta_N/q_{95}^2$) as function of discharge duration, closed symbols are transient discharges, closed symbols are stationary, the duration ($W>0.85$ maximum $W$) is normalised to the energy confinement time ($\tau_E$) averaged during this time window. (b) $H_{89}\beta_N/q_{95}^2$ versus $\varepsilon^{0.5}\beta_p$. The lines indicates the different values for $q_{95}$, transient (open symbols) and stationary results (closed symbols) are given.

**Figure 10:**     Normalised beta, $\beta_N$, as function of the pressure peaking, $p_0/<p>$, for different advanced regimes: (a) reversed shear discharges and (b) hybrid discharges. Transient (open symbols) and stationary results (closed symbols) are given. Discharges from various devices are colour coded (see legend).

**Figure 11:**     Plasma cross sections for four conceptual models (A,B,C and D) from the European Power Plant Conceptual Study [60], compared to ITER.

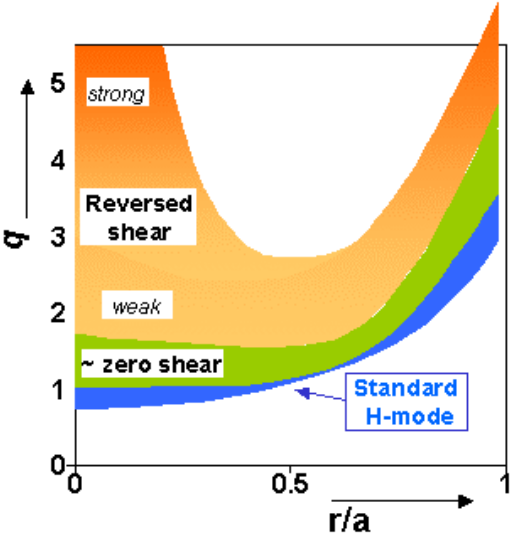

Figure 1

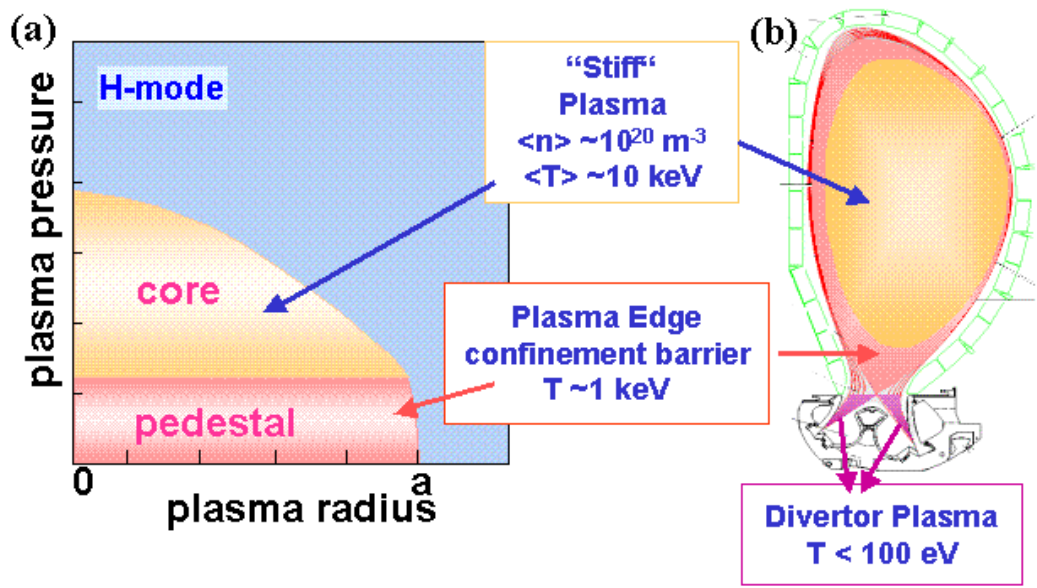

Figure 2

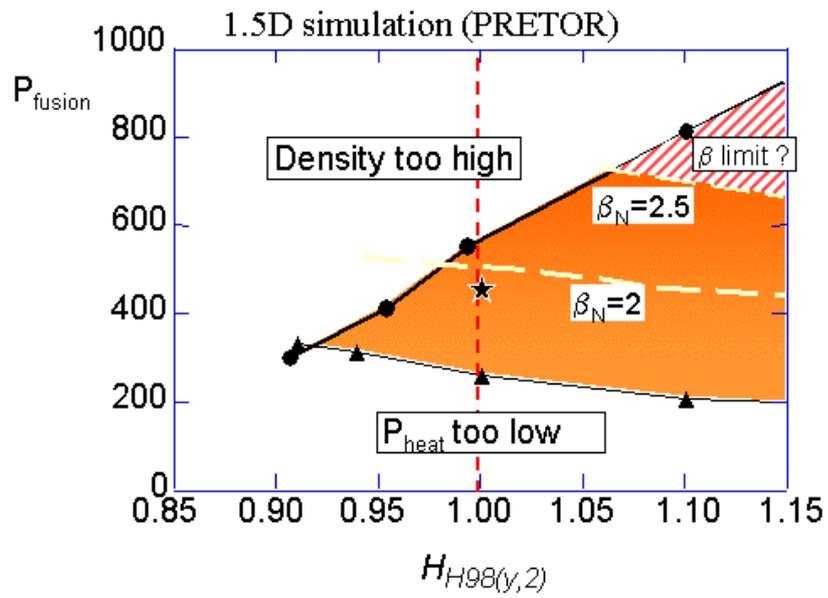

Figure 3

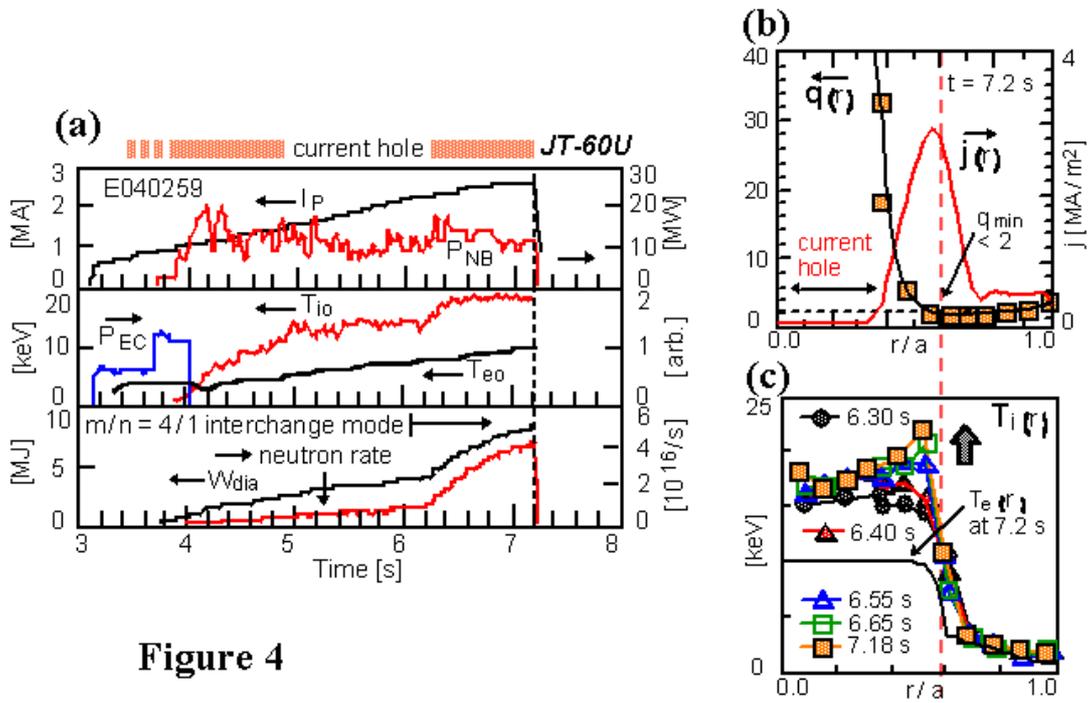

Figure 4

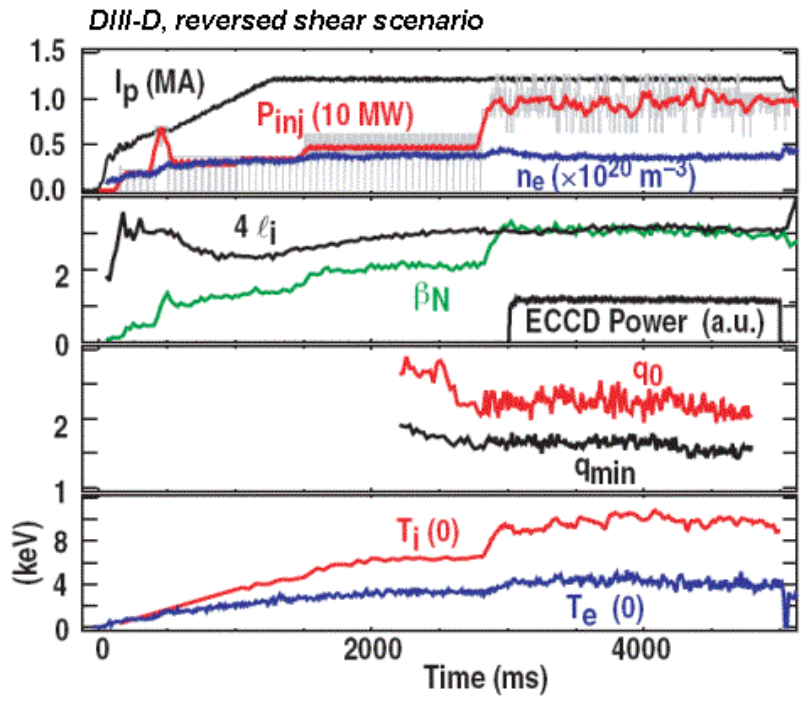

Figure 5

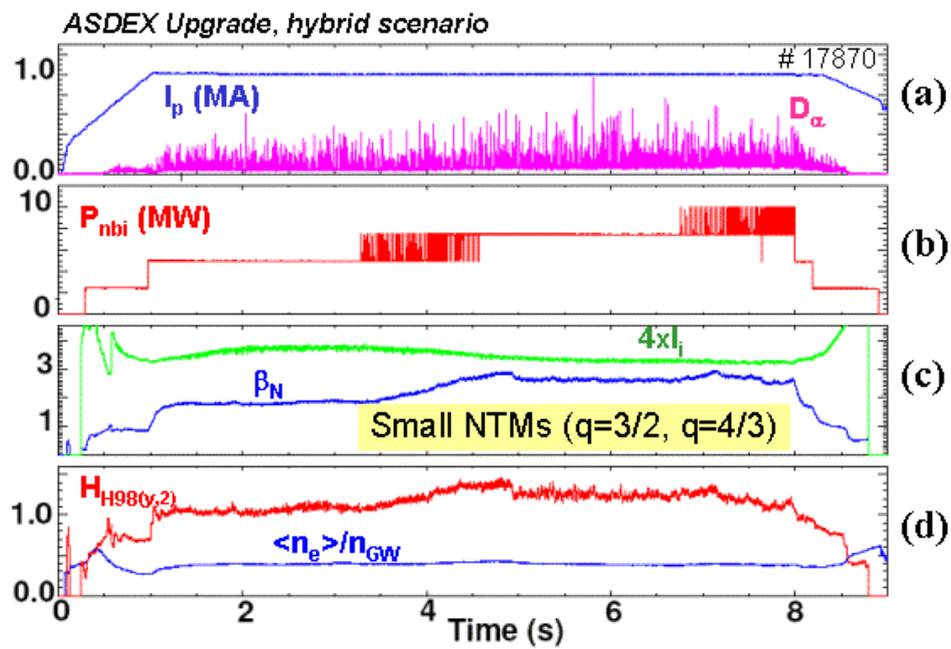

Figure 6

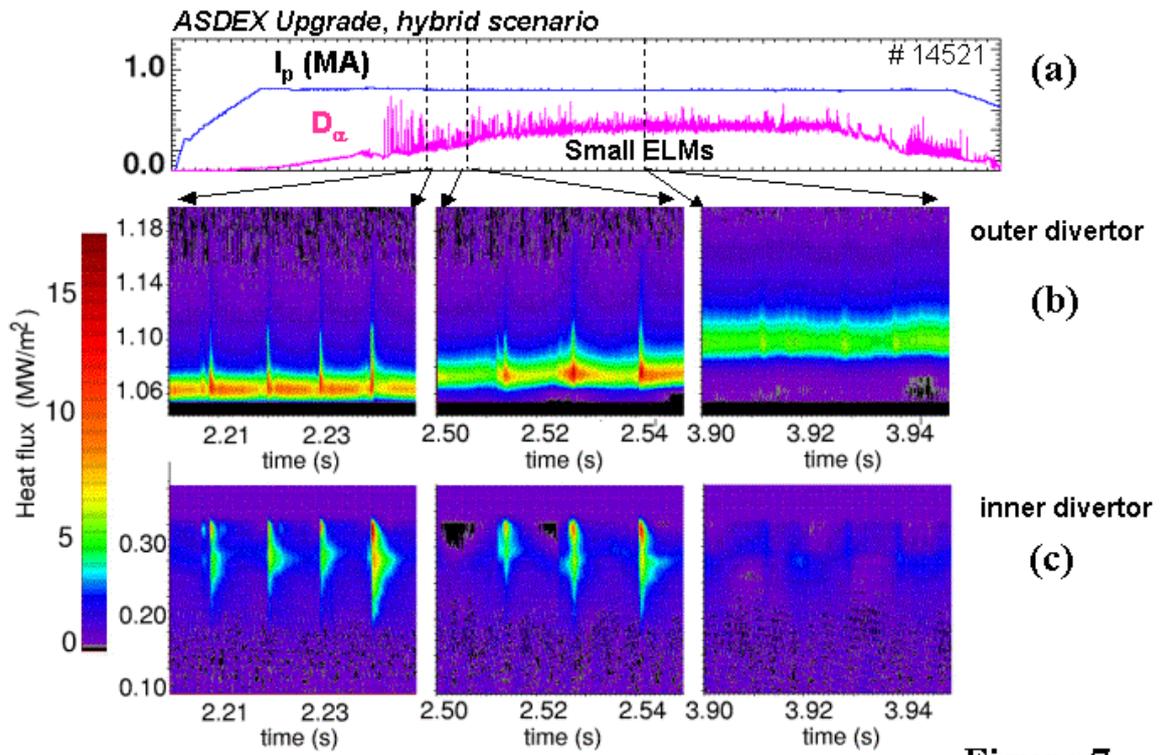

Figure 7

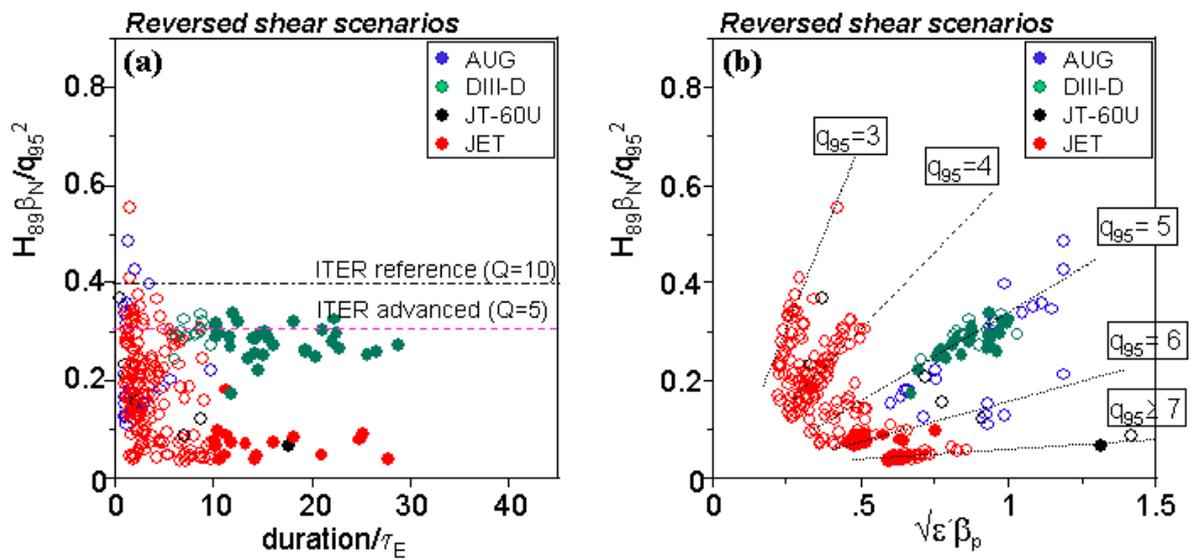

Figure 8

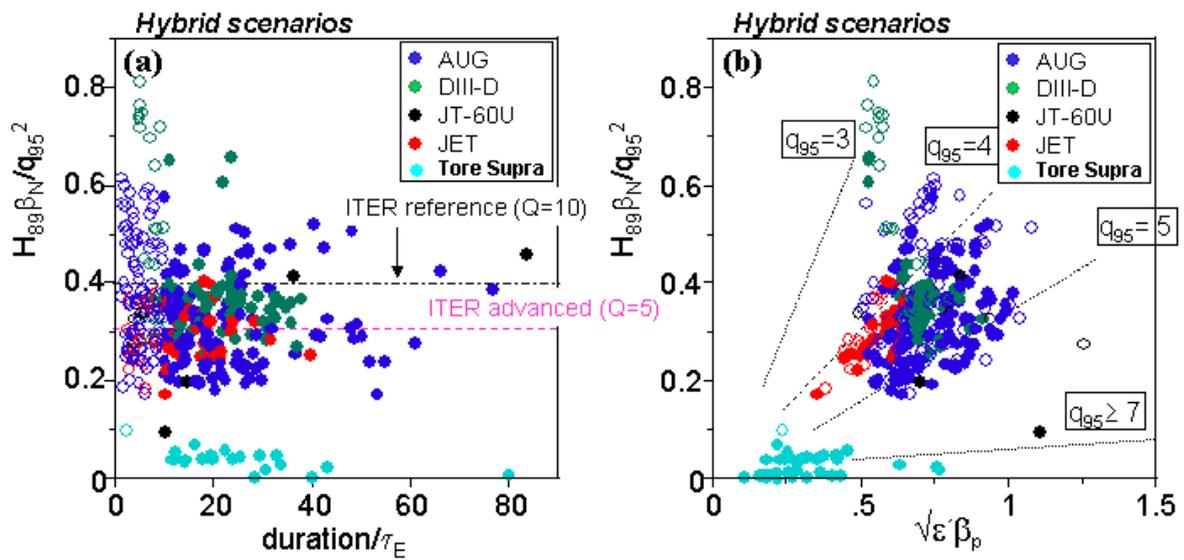

Figure 9

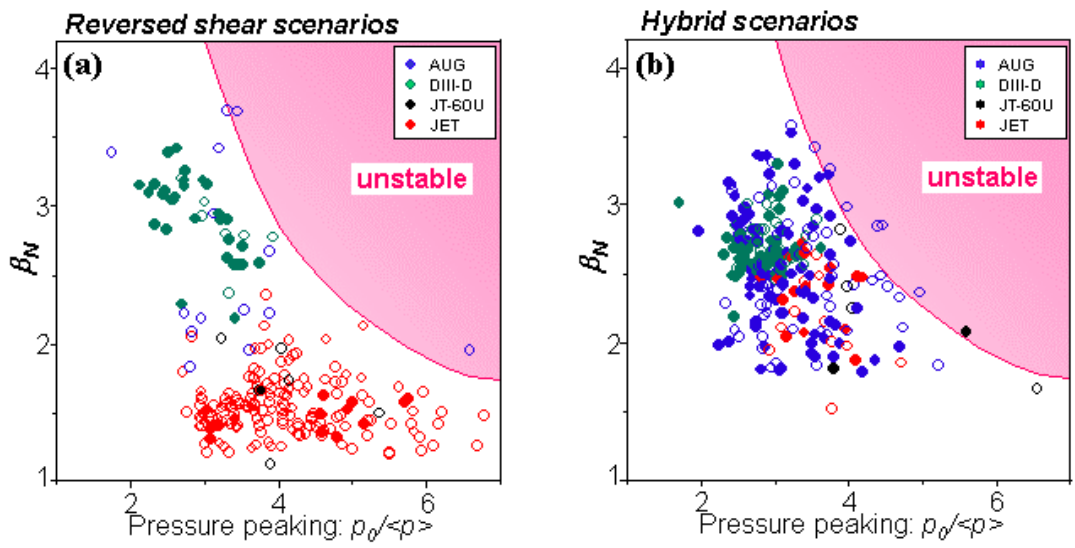

Figure 10

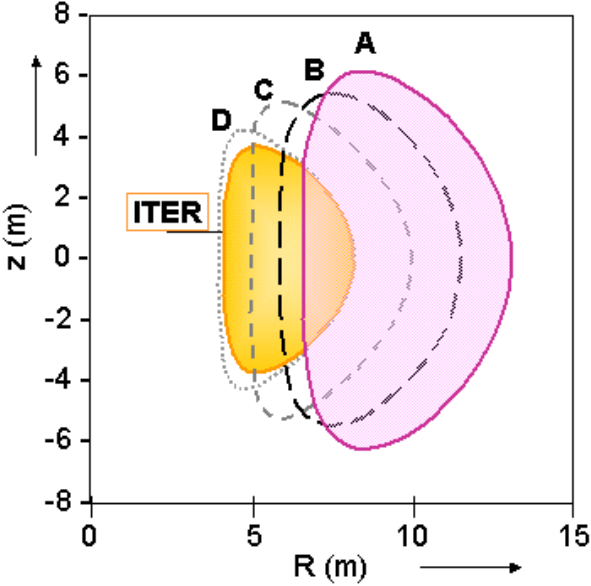

Figure 11